\documentclass[aps,pra,twocolumn,showpacs]{revtex4-1}
\usepackage{indentfirst}
\usepackage{bm}
\usepackage{verbatim}
\usepackage{graphicx}
\usepackage{subfigure}
\usepackage{epsfig}
\usepackage{amsmath}
\usepackage{booktabs} 
\usepackage{multirow}
\usepackage{txfonts} 
\usepackage{float}
\usepackage{tabularx}
\usepackage{threeparttable}
\usepackage{appendix}
\usepackage[citecolor=blue,colorlinks=true,linkcolor=blue,urlcolor=blue,breaklinks=true]{hyperref}

\begin{document}
\title{Theory of anomalous Landau-Zener tunneling induced by nonlinear coupling }%
\author{Wen-Yuan Wang}
\email[E-mail address: ]{wywang@nwnu.edu.cn.}
\author{Hong-Juan Meng}
\email[E-mail address: ]{mhjnx@163.com.}
\affiliation{Key Laboratory of Atomic and Molecular Physics $\&$ Functional Materials of Gansu Province, College of Physics and Electronic Engineering, Northwest Normal University, Lanzhou 730070, China 
}
\begin{abstract}
We develop a general theory of Landau-Zener (LZ) tunneling in a two-level system with amplitude-dependent, sign-reversible nonlinear coupling, distinguishing it fundamentally from conventional on-site nonlinearity. Through a combination of analytical and phase-space analysis, we show that beyond a critical interaction strength, the nonlinear coupling fundamentally reshapes the adiabatic energy landscape, introducing a topological twisted and knotted structure. This structure leads to a complete breakdown of the standard exponential LZ formula, even in the adiabatic limit. Central to this anomalous behavior is the emergence of a black-hole-like fixed point, which acts as a universal attractor: upon traversing the critical region, all quantum trajectories converge to this fixed point, irreversibly erasing any memory of the initial state. From this fixed-point picture, we derive an exact analytical expression for the adiabatic tunneling probability, revealing a characteristic power-law dependence on both linear and nonlinear coupling strength. Our work establishes a paradigmatic framework for nonlinear-coupling-induced anomalous adiabaticity breaking and offers a universal mechanism for state control in driven quantum and wave systems.
\end{abstract}

\maketitle
\section{Introduction}
The Landau-Zener (LZ) model is one of cornerstone of quantum dynamics, describing the nonadiabatic transition between two energy levels under a linearly varying external field  \cite{Landau1932,Zener1932}. Its celebrated exponential tunneling formula has been validated across diverse physical platforms, from superconducting qubits \cite{PhysRevLett.96.187002,PhysRevA.74.052330,Berns2008,IVAKHNENKO20231}, nitrogen-vacancy centers \cite{doi:10.1126/science.1181193}, quantum dots \cite{PhysRevLett.102.216802,doi:10.1126/science.1183628}, waveguide arrays \cite{PhysRevLett.94.113904,PhysRevLett.126.054301}, and Bose-Einstein condensates \cite{PhysRevA.61.023402,PhysRevA.61.033603,PhysRevA.66.023404,PhysRevA.66.063603,PhysRevLett.87.140402,PhysRevA.65.063612,
PhysRevLett.90.170404,PhysRevLett.91.230406,PhysRevA.73.063609,PhysRevLett.96.020405,PhysRevLett.102.230401,Zenesini_2008,
PhysRevLett.103.090403,Chen2011,PhysRevLett.106.155302,Zenesini_2008,PhysRevLett.103.090403,Chen2011,
PhysRevLett.120.040407,PhysRevA.99.023616,PhysRevLett.125.213401}, to name only a few. The LZ model has been to nonlinear regimes where the level energies depend on the state occupation \cite{PhysRevA.61.023402,PhysRevA.61.033603,PhysRevA.66.023404,PhysRevA.66.063603,PhysRevLett.87.140402,PhysRevA.65.063612,
PhysRevLett.90.170404,PhysRevLett.91.230406,PhysRevA.73.063609,PhysRevLett.96.020405,PhysRevLett.102.230401,Zenesini_2008,
PhysRevLett.103.090403,Chen2011,PhysRevLett.106.155302,Zenesini_2008,PhysRevLett.103.090403,Chen2011,
PhysRevLett.120.040407,PhysRevA.99.023616,PhysRevLett.125.213401,PhysRevA.106.062613,PhysRevA.107.032420,PhysRevA.110.043314}---a phenomenon arising from mean-field interactions in Bose--Einstein condensates, Josephson systems, and nonlinear photonic structures. Recently, the extensions of the LZ tunneling to non-Hermitian systems have  also been studied by considering  level decay and dephasing effect \cite{PhysRevB.36.2770,PhysRevA.46.4110,PhysRevA.90.032116,PhysRevA.104.013111,PhysRevA.109.012622}, as well as nonreciprocal coupling \cite{PhysRevA.106.063708,PhysRevA.108.063506,PhysRevA.108.062217}.

However, existing studies of the nonlinear LZ problem have predominantly focused on on-site nonlinearity (e.g., Kerr-type terms), where the interaction depends locally on the density at each site \cite{PhysRevA.61.023402,PhysRevA.61.033603,PhysRevA.66.023404,PhysRevA.66.063603,PhysRevLett.87.140402,PhysRevA.65.063612,
PhysRevLett.90.170404,PhysRevLett.91.230406,PhysRevA.73.063609,PhysRevLett.96.020405,PhysRevLett.102.230401,Zenesini_2008,
PhysRevLett.103.090403,Chen2011,PhysRevLett.106.155302,Zenesini_2008,PhysRevLett.103.090403,Chen2011,PhysRevLett.120.040407,
PhysRevA.99.023616,PhysRevLett.125.213401,PhysRevA.106.062613,PhysRevA.107.032420,PhysRevA.110.043314}. In such systems, the energy spectrum develops looped structures and exhibits hysteresis. For subcritical interaction strengths, the tunneling probability, though modified, often retains an exponential form in the adiabatic limit. A key finding is the pronounced breakdown of adiabaticity for sufficiently large nonlinear parameters, a phenomenon directly linked to the hysteresis and the swallowtop loops in the spectrum \cite{PhysRevA.61.023402, PhysRevA.66.023404, PhysRevA.66.063603, Eckel2014}. The underlying mechanism has been effectively explained by mapping the system to an equivalent classical Josephson Hamiltonian, where a nonzero tunneling probability in the adiabatic limit corresponds to a jump in the classical canonical action \cite{PhysRevA.66.023404,PhysRevLett.90.170404,PhysRevA.106.063708}.
In contrast, a fundamentally distinct and less explored class of nonlinearity for the nonlinear LZ problem arises from nonlinear coupling, where the interaction between subsystems depends on their relative amplitudes or phases \cite{PhysRevB.93.155112,PhysRevLett.123.053902,Hadad2018,doi:10.1126/science.abd2033,Zhou2022,Sone2024,Sone2025,2505.09179,6bsr-x2v9}. This form of nonlinearity opens a different dynamical regime. Recent experiments in acoustic and circuit systems have demonstrated that nonlinear coupling can engineer unusual multistability and non-trivial state-transfer pathways, suggesting a rich, untapped dynamical regime \cite{6bsr-x2v9}.

In this paper, we introduce and solve a minimal LZ model with programmable nonlinear coupling. Contrary to on-site nonlinearity, we show that nonlinear coupling reshapes the adiabatic energy landscape into a topologically twisted structure that supports a continuous line of fixed points in phase space. This structure acts as a ¡°black-hole-like¡± attractor, fundamentally altering the adiabatic flow and leading to a complete breakdown of the exponential tunneling law. We derive an exact analytical expression for the tunneling probability, revealing a universal power-law scaling near the critical point. Our theory not only explains recent observations of anomalous state transfer in coupled resonator systems but also provides a general framework for designing and controlling nonadiabatic transitions in quantum and classical wave platforms.

This paper is organized as follows. In Sec.~\ref{model}, we introduce the physical model of the LZ system with nonlinear coupling. Section~\ref{AdiabaticEnergy} analyzes how nonlinear coupling fundamentally reshapes the adiabatic energy landscape and elucidates its role in driving phase transitions. In Sec.~\ref{LZTunneling}, we investigate the anomalous LZ tunneling induced by nonlinear coupling, with particular emphasis on the breakdown of adiabatic following. A new class of nonlinear quantum adiabatic theorem is developed in Sec.~\ref{AdiabaticTheorem}. Finally, Sec.~\ref{Conclusion} provides a comprehensive discussion and concludes the paper.

\section{Landau-Zener model with nonlinear coupling}\label{model}
We consider a minimal two-level system with amplitude-dependent coupling, described by the nonlinear Schr\"{o}dinger equation (\(\hbar=1\)):
\begin{equation}\label{eq:ham}
i\frac{d}{dt}\begin{pmatrix} a \\ b \end{pmatrix} = H(\gamma)\begin{pmatrix} a \\ b \end{pmatrix}, \quad
H(\gamma)=\left(\begin{array}{cc}
\gamma &   \alpha+\beta|a|^{2} \\
\alpha+\beta|a|^{2} &   -\gamma
\end{array}\right).
\end{equation}
Here, \(a\) and \(b\) are probability amplitudes (\(|a|^2+|b|^2=1\)), \(\gamma(t)=v t\) is the linearly swept bias with $v$ is the sweeping rate, $\alpha>0$ is the linear coupling, and $\beta$ introduces an amplitude-dependent sign-reversible coupling coupling. This form of coupling coupling directly mirrors recent experimental realizations in acoustic dimers and active metamaterials \cite{6bsr-x2v9}. Since the Hamiltonian can be scaled by dividing by $\alpha$, for convenience, we can set $\alpha=1$ as the energy unit hereafter.

\section{Nonlinear coupling fundamentally reshapes the adiabatic energy landscape}\label{AdiabaticEnergy}
we first analyze the behavior of the adiabatic levels in the nonlinear model in order to understand the LZ tunneling for the new problem. The adiabatic eigenvalues \(\varepsilon(\gamma)\) are obtained by solving the nonlinear eigenvalue problem \(H(\gamma)\psi = \varepsilon\psi\). For \(\beta=0\), we recover the standard avoided crossing with gap \(\Omega=2\alpha\). For \(\beta \neq 0\), the equation for the intensity \(I=|a|^2\) becomes a higher-order polynomial. We find the eigenenergy $\varepsilon$ satisfies the following quartic equation:
\begin{equation}\label{eq:nonliearlevel}
\varepsilon^4+D\varepsilon^2+E\varepsilon+F=0 \;,
\end{equation}
\mbox{where $D=-\frac{1}{4} \left(4 \alpha ^2+4 \alpha  \beta +\beta ^2+4 \gamma ^2\right)$, $E=-\frac{1}{4} \left(4 \alpha  \beta  \gamma +2 \beta ^2 \gamma \right)$}, and $F=-\frac{1}{4} \beta ^2 \gamma ^2$. When $\delta =0$, the above equation reduces to that of \cite{PhysRevA.61.023402,PhysRevA.106.063708}.

\begin{figure}[tb]
\centering
\includegraphics[width=\columnwidth]{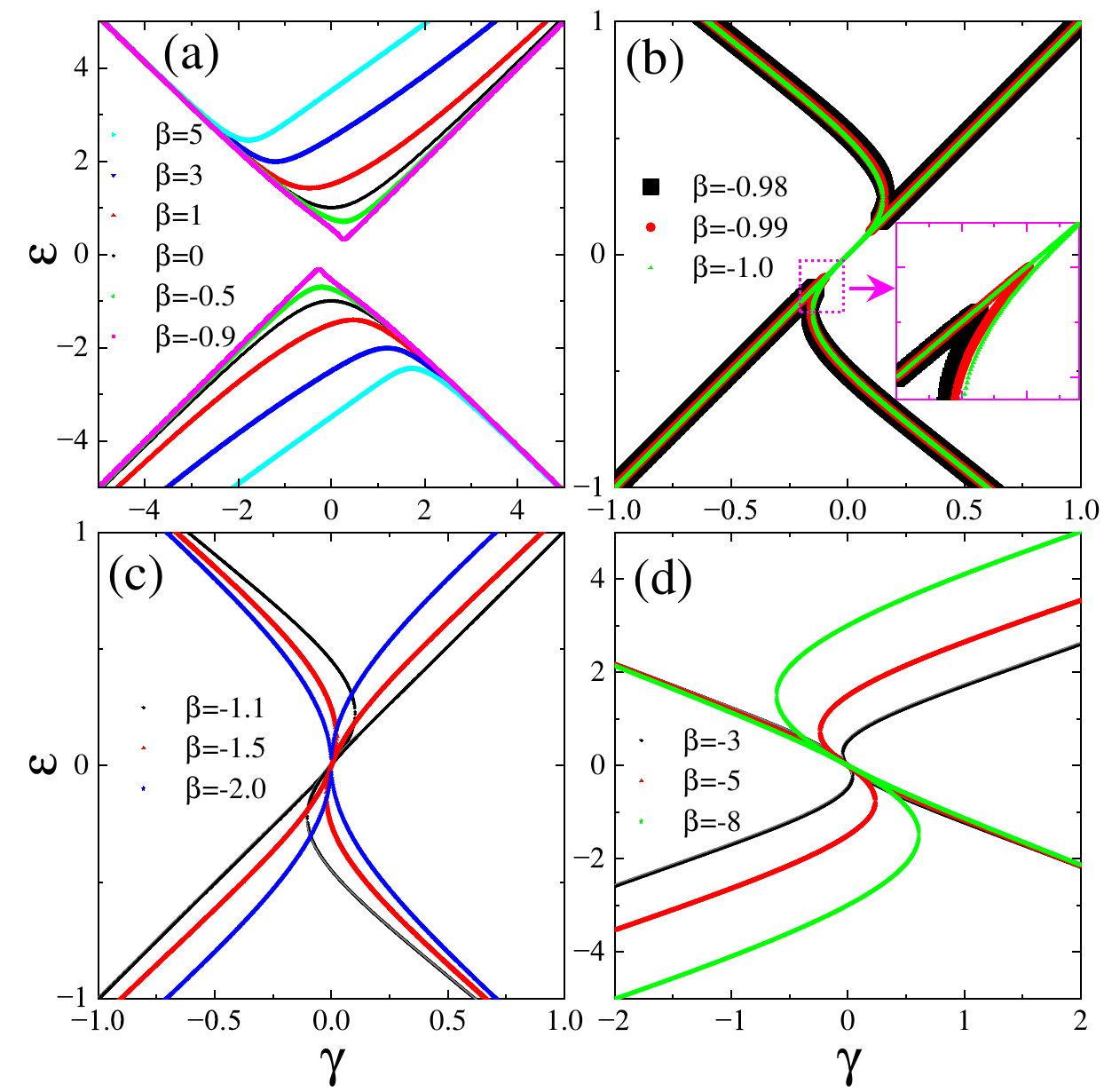}
\caption{(color online) Adiabatic energy levels $\varepsilon$ as a function of bias $\gamma$ for different regimes of the nonlinear coupling strength $\beta$. The nonlinear coupling qualitatively alters the level structure and induces four distinct topological types: (a) Type-I, a conventional avoided crossing for $\beta\gtrsim-0.9575$; (b) Type-II, a swallowtail structure for $-1<\beta\lesssim-0.9575$; (c) Type-III, a twisted-knotted structure A for $-2<\beta<-1$; and (d) Type-IV, a twisted-knotted structure B for $\beta<-2$. }
\label{fig:energylevels}
\end{figure}

Depending on the parameters, this quartic equation may admit either two or four real roots. In Fig. \ref{fig:energylevels}(a), for a nonlinear coupling strength $\beta \gtrsim -0.9575$, the level structure exhibits a conventional avoided crossing. In the absence of nonlinear coupling ($\beta = 0$), the upper and lower energy levels are symmetric about $\gamma = 0$, with a minimum gap $\Omega = 2\alpha$ at the symmetry point. In this regime, as the nonlinear coupling strength increases, the gap between the two levels widens, and the symmetry about $\gamma = 0$ is broken. The asymmetry becomes more pronounced with increasing nonlinear coupling strength. We refer to this level structure as a conventional avoided crossing (Type-I).

A swallowtail structure emerges in the adiabatic spectrum when the nonlinear coupling strength lies in the transitional regime $-1 < \beta \lesssim -0.9575$, as shown in Fig.~\ref{fig:energylevels}(b). Within the swallowtail region, the number of real energy levels increases to four. The swallowtail becomes increasingly pronounced as the nonlinear coupling strength grows. We classify this pattern as a swallowtail structure (Type-II).

When the nonlinear coupling strength exceeds the critical value \(\beta = -1\), the upper and lower adiabatic levels connect at \(\gamma = 0\), forming a twisted--knotted structure [Figs.~\ref{fig:energylevels}(c) and \ref{fig:energylevels}(d)]. For a given \(\beta\) in this regime, a corresponding threshold \(\gamma_c\) exists such that four real energy levels coexist within the interval \(-\gamma_c < \gamma < \gamma_c\). The value of \(\gamma_c\) varies piecewise with \(|\beta|\):
\begin{itemize}
    \item For \(-2 < \beta < -1\), \(\gamma_c\) decreases from a maximum of \(\approx0.1843\) at \(\beta = -1\) to a minimum of \(0\) at \(\beta = -2\).
    \item For \(\beta < -2\), \(\gamma_c\) increases with growing \(|\beta|\); hence the region supporting four real roots expands.
\end{itemize}
Moreover, the slope of the energy levels at the knot point exhibits a piecewise dependence on \(\beta\):
\begin{itemize}
    \item In the range \(-2 < \beta < -1\), the slope is positive. It rises rapidly from \(1\) near \(\beta = -1\) and diverges as \(\beta \rightarrow -2\).
    \item For \(\beta < -2\), the slope becomes negative. Starting from \(-\infty\) at \(\beta = -2\), its magnitude gradually decreases and asymptotically approaches \(-1\) for strongly negative \(\beta\).
\end{itemize}
To distinguish these two regimes, we refer to the structure in the interval \(-2 < \beta < -1\) as twisted--knotted structure A (Type-III), and that for \(\beta < -2\) as twisted--knotted structure B (Type-IV).
\begin{figure}[tb]
\centering
\includegraphics[width=\columnwidth]{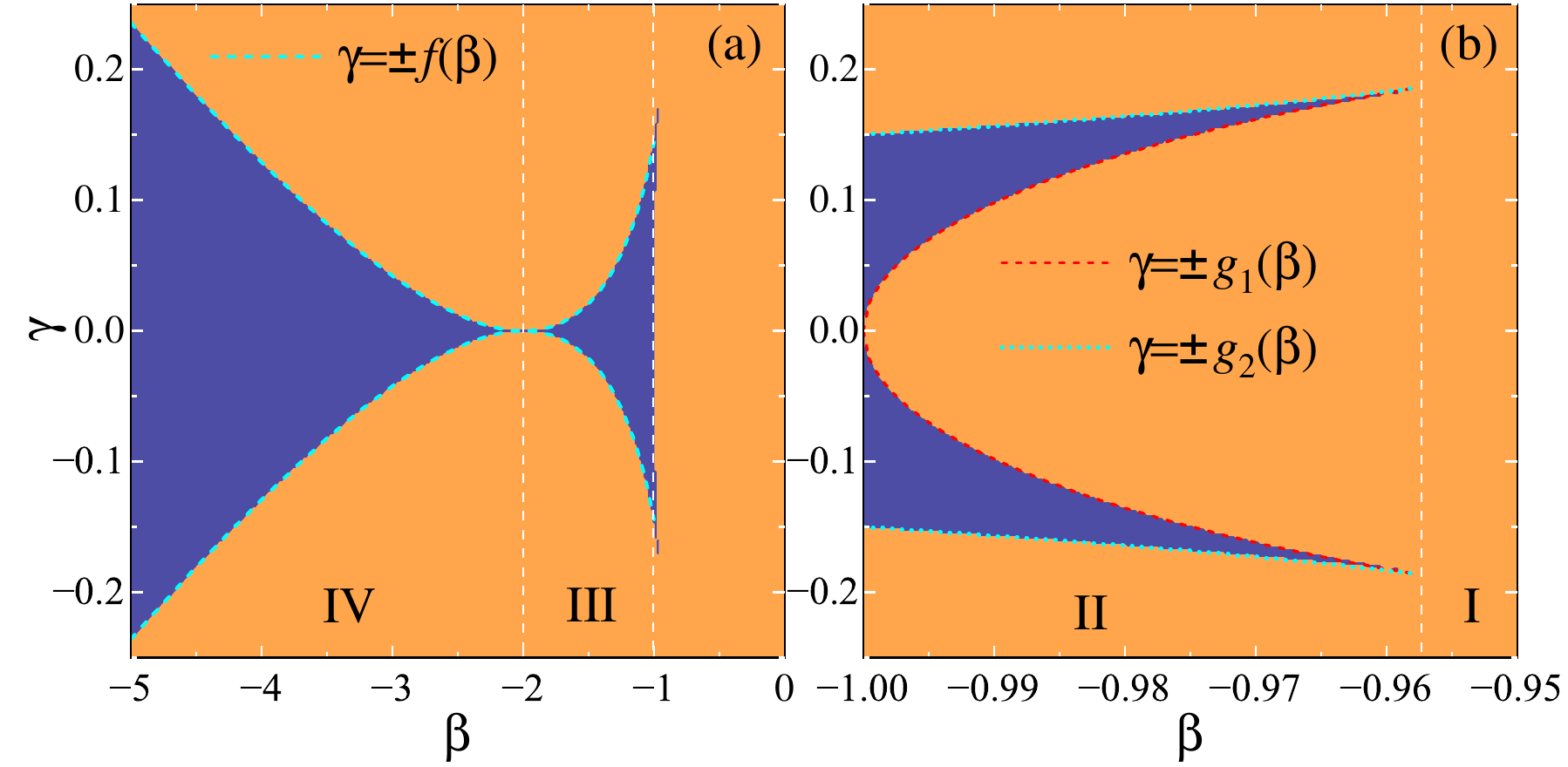}
\caption{(color online) Phase diagram in the ($\beta,~\gamma$) parameter space, showing regions with two real adiabatic energy levels (orange) and four real adiabatic energy levels (shaded blue). The analytical boundaries $f(\beta)$, $g_1(\beta)$, and $g_2(\beta)$, derived from the fixed-point analysis described in the text (Eqs. \eqref{eq:boundary1} and \eqref{eq:boundary2}), are indicated and show excellent agreement with the numerically determined shaded regions. The labels I, II, III, and IV correspond to the four distinct topological types illustrated in Fig.~\ref{fig:energylevels}. }
\label{fig:energyphase}
\end{figure}
To characterize the parameter regimes in which the adiabatic spectrum possesses either two or four real energy levels, we construct a phase diagram in the plane of the nonlinear coupling strength \(\beta\) and the bias \(\gamma\). As shown in Fig.~\ref{fig:energyphase}, the blue shaded region corresponds to combinations of \(\beta\) and \(\gamma\) for which four real adiabatic levels emerge, whereas the unshaded (orange) region indicates where only two real levels exist. A magnified view of the critical region near \(\beta \sim -1\) is presented in Fig.~\ref{fig:energyphase}(b).

The phase diagram is partitioned according to the four distinct topological types identified in Fig.~\ref{fig:energylevels}, with the corresponding labels I-IV marked in their respective parameter domains. One can clearly observe that in Type-IV (\(\beta < -2\)), the region supporting four real levels expands as both \(|\beta|\) and \(|\gamma|\) increase. In contrast, for Type-III (\(-2 < \beta < -1\)), the four-level region is confined within a narrow window \(|\gamma|<0.25\). Type-II represents a very narrow transitional regime (\(-1 < \beta \lesssim -0.9575\)) in which a swallowtail forms only over a small interval of \(\beta\). For \(\beta \lesssim -0.9575\) (Type-I), only two real levels exist for any value of \(\gamma\), corresponding to the conventional avoided-crossing structure.

The numerically computed boundary between the two- and four-level regions appears smooth and well defined. In the following section, we derive explicit analytical expressions for this boundary through a fixed-point analysis. The resulting curves, denoted \(f(\beta)=(\beta+2)^2/(8\beta\sqrt{1-(0.5+1/\beta)^2})\), \(g_1(\beta)= \sqrt{\kappa} \left( 1 - 2\kappa - 6\kappa^2 \right) \), and \(g_2(\beta)=0.15+0.636\kappa+2.4\kappa^2+40\kappa^3\), with $\kappa=\beta + 1$ are plotted in Fig.~\ref{fig:energyphase} and show perfect agreement with the numerical phase boundaries.

\section{Nonlinear coupling induced anomalous LZ tunneling}\label{LZTunneling}
\subsection{Landau-Zener tunneling induced by nonlinear coupling}
The emergence of swallowtail and twisted-knotted structures in the adiabatic spectrum leads to a remarkable phenomenon: the breakdown of adiabatic evolution even in the adiabatic limit. As predicted in nonlinear on-site interaction extensions of the LZ model \cite{PhysRevA.61.023402,PhysRevA.61.033603,PhysRevA.66.023404}, when the nonlinear parameter exceeds a critical value, the system can no longer follow the instantaneous eigenstate adiabatically, resulting in a finite tunneling probability regardless of how slowly the bias is swept.

\begin{figure}[tb]
\centering
\includegraphics[width=\columnwidth]{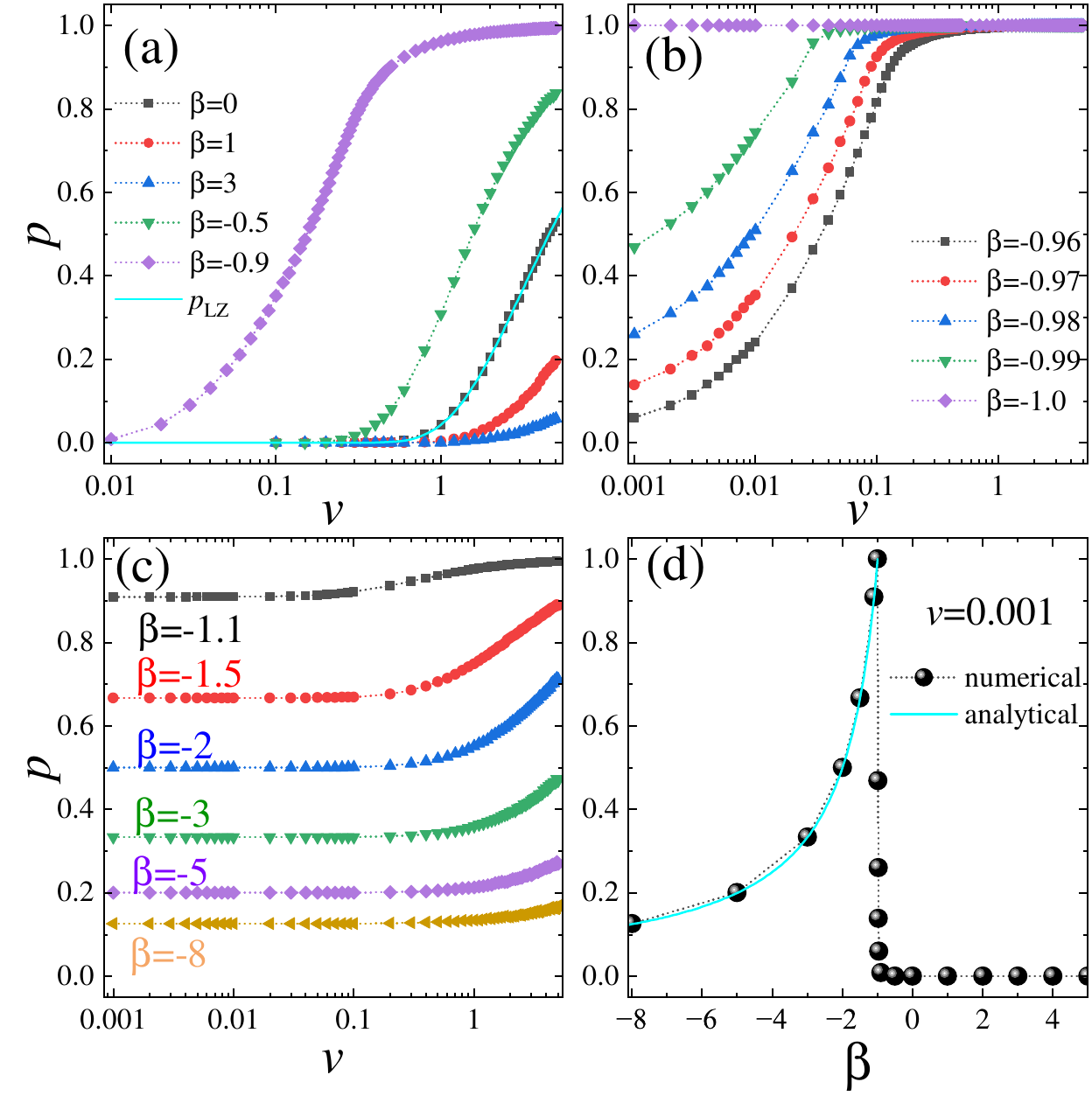}
\caption{(color online) Landau-Zener tunneling probability \(p\) induced by nonlinear coupling $\beta$, plotted as a function of the sweep rate \(v\) for (a) the conventional avoided-crossing level structure regime (Type-I), (b) the swallowtail level structure regime (Type-II), and (c) the twisted-knotted level structure regime (Type-III and Type-IV). For comparison, panel (a) includes the standard LZ result for the linear case (\(\beta = 0\)). Panel (d) shows the adiabatic tunneling probability as a function of nonlinear coupling \(\beta\) with \(v = 0.001\); numerical data are represented by solid symbols, while the solid line corresponds to the analytical expression Eq. \eqref{eq:pad} obtained from fixed-point analysis.}
\label{fig:fig3}
\end{figure}

To examine this effect in our model with nonlinear coupling, we initialize the system in the lower adiabatic branch, corresponding to \((a,~b)^\mathrm{T}=(1,~0)\) as \(\gamma \to -\infty\). The bias \(\gamma\) is then increased linearly. Upon passing through the swallowtail or twisted-knotted region, nonadiabatic transitions to the upper branch occur. The tunneling probability, defined as the population transferred to the upper level after the sweep, depends sensitively on the nonlinear coupling strength \(\beta\) and the sweep rate \(v\).

Figure \ref{fig:fig3} presents the tunneling probability across the different regimes of \(\beta\). Panel (a) shows the tunneling probability \(p\) as a function of the sweep rate \(v\) for the conventional avoided-crossing structure regime (Type-I, \(\beta >-0.9575\)). In this regime, the probability follows a modified exponential law. For reference, the standard LZ result \(p_{\mathrm{LZ}}=e^{-\pi\alpha^2/v}\) (corresponding to \(\beta = 0\)) is included; as expected, the numerical data coincide with this analytic curve when \(\beta = 0\). Interestingly, in the fast-sweep case, the non-adiabatic tunneling probability decreases as the value of \(\beta\) increases from -0.9 to 3. This reduction is directly correlated with the widening of the avoided-crossing gap shown in Fig. \ref{fig:energylevels}(a), since a larger gap suppresses LZ transitions for a given sweep rate, a trend consistent with the physical picture underlying the standard LZ formula.

When the quantum-state evolution takes place in the regime of a swallowtail structure (Type-II, \(-1 < \beta \lesssim -0.9575\)), as illustrated in Fig. \ref{fig:fig3}(b), a finite adiabatic tunneling probability emerges even in the limit of an infinitely slow sweep. This breakdown of perfect adiabatic following originates fundamentally from the topological reorganization of the energy levels---specifically, the appearance of the swallowtail---and is qualitatively analogous to the effect induced by on-site nonlinearity in earlier studies \cite{PhysRevA.61.023402,PhysRevA.61.033603,PhysRevA.66.023404}.
Notably, the adiabatic tunneling probability increases as the nonlinear coupling strength grows (i.e., as \(\beta\) becomes more negative). This trend can be directly traced to the evolution of the level structure shown in Fig. \ref{fig:energylevels}(b): with stronger nonlinear coupling, the two swallowtails associated with the upper and lower branches move closer together, effectively reducing the energetic barrier that separates them and thereby enhancing the adiabatic transition. At the critical value \(\beta = -1\), the swallowtails connect at \(\gamma = 0\), leading to complete adiabatic tunneling. Equally important, in the fast-sweep (non-adiabatic) regime, the same tendency of the swallowtails to approach each other results in a high tunneling probability even for rapid sweeps. This behavior suggests that nonlinear coupling provides a viable route for efficient non-adiabatic control of quantum states, offering a complementary strategy to purely adiabatic protocols.

Most strikingly, when the quantum state evolves through the twisted-knotted level structure regimes (Type-III and Type-IV), one might intuitively expect complete state transfer (tunneling probability of 1) via the knot point. However, our calculations reveal a distinctly different picture, as displayed in Fig. \ref{fig:fig3}(c). Both Type-III (\(-2 < \beta < -1\)) and Type-IV (\(\beta < -2\)) exhibit a nonzero adiabatic tunneling probability in the limit of vanishing sweep rate. This probability decreases as the nonlinear coupling strength grows (i.e., as \(\beta\) becomes more negative), falling from unity near \(\beta \approx -1\). In this regime, the tunneling probability shows a relatively robust dependence on the sweep rate, in contrast to the conventional exponential behavior.

To clarify the relationship between adiabatic tunneling and nonlinear coupling, Fig. \ref{fig:fig3}(d) presents the adiabatic tunneling probability as a function of \(\beta\) for a fixed slow sweep rate \(v = 0.001\). The data clearly distinguish the three characteristic regimes: In Type-I regime, the finite gap suppresses adiabatic transitions entirely, yielding zero adiabatic tunneling probability. In the narrow transitional Type-II regime, the adiabatic tunneling probability depends sensitively and non-monotonically on \(\beta\). Most notably, both in the Type-III and Type-IV regimes---where the energy levels are connected at the knot---a finite adiabatic tunneling probability persists, contradicting the naive expectation of perfect adiabatic passage. This unexpected residual tunneling originates from the topological rearrangement of the phase-space fixed points, which we analyze in detail below. The solid line in Fig. \ref{fig:fig3}(d) represents the analytical result derived from fixed-point analysis [Eq. \eqref{eq:pad}], showing excellent agreement with the numerical data points and confirming the reliability of our theoretical description.

\subsection{Nonlinear coupling induced the breakdown of adiabatic following}
\begin{figure}[tb]
\centering
\includegraphics[width=\columnwidth]{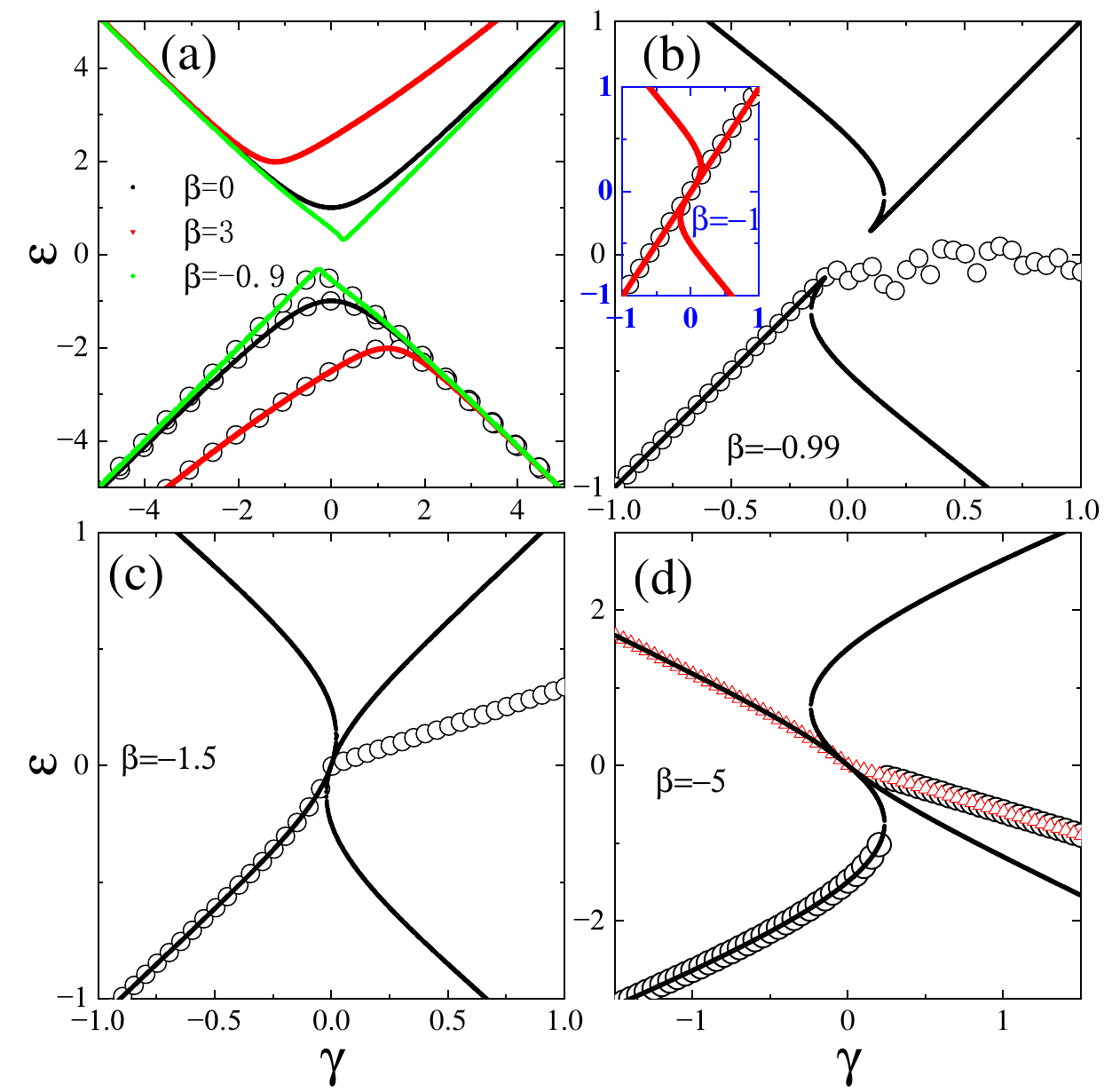}
\caption{(color online) Nonlinear coupling induced adiabatic following versus breakdown in quantum-state evolution. Comparison between the dynamical energy levels \(\varepsilon_{\mathrm{dyn}}\) (symbols) and the adiabatic levels (solid lines) in (a) Type-I regime, (b) Type-II regime, (c) Type-III regime, and (d) Type-IV regime. In (a)-(d), the dynamical levels (circles) are obtained by evolving from the lower adiabatic branch. In (d), the dynamical levels (triangles) are evolved from the upper adiabatic branch to illustrate the distinct behavior in this regime. All dynamical evolutions are computed in the adiabatic limit with a sweep rate \(\alpha = 0.001\). }
\label{fig:fig4}
\end{figure}
To gain insight into how nonlinear coupling governs adiabatic following versus its breakdown in quantum-state evolution, we compute the dynamical energy along the temporal evolution \cite{PhysRevA.61.023402,PhysRevLett.125.213401}. For a wave function \(\psi(\gamma=\alpha t) = \big(a(t), b(t)\big)^{\!\mathrm T}\) that satisfies the Schr\"{o}dinger equation with the Hamiltonian (\ref{eq:ham}), the dynamical energy at a given bias \(\gamma\) is defined as the instantaneous expectation value
\begin{equation}
\varepsilon_{\mathrm{dyn}}(\gamma)=\big\langle\psi(\gamma)\big|H(\gamma)\big|\psi(\gamma)\big\rangle .
\label{eq:dynE}
\end{equation}
According to the adiabatic theorem, if the sweep is sufficiently slow, the dynamical energy should closely track one of the adiabatic eigenvalues. A persistent deviation of \(\varepsilon_{\mathrm{dyn}}\) from the adiabatic branch, even in the limit \(\alpha\to 0\), signals a breakdown of adiabaticity and implies a finite tunneling probability between the adiabatic states \cite{PhysRevA.61.023402,PhysRevLett.125.213401}.

Figure \ref{fig:fig4} compares the dynamical levels (symbols) with the adiabatic levels (solid lines) for the four characteristic regimes. In the Type-I regime [Fig. \ref{fig:fig4}(a)], the dynamical level follows the lower adiabatic branch almost perfectly, mirroring the behavior of the linear case (\(\beta=0\)) \cite{PhysRevA.61.023402}. This close tracking is a direct consequence of the conventional avoided-crossing structure and the finite gap that separates the two adiabatic states; here the nonlinear coupling merely renormalizes the gap without spoiling adiabaticity. Hence, in this regime nonlinear coupling can be used to protect adiabatic evolution by widening the gap and further suppressing non-adiabatic transitions.

The situation changes qualitatively in the Type-II regime [Fig. \ref{fig:fig4}(b)]. While the dynamical level initially follows the lower adiabatic branch, it clearly departs from that branch upon entering the swallowtail region. This departure reflects the onset of adiabatic breakdown, analogous to that induced by a large on-site nonlinearity in earlier studies \cite{PhysRevA.61.023402,PhysRevA.66.023404,PhysRevA.66.063603,PhysRevLett.90.170404,Eckel2014}. Because the Type-II regime occupies only a very narrow window in parameter space (\(-1<\beta\lesssim-0.9575\)) and its behavior is well understood from earlier work on swallowtail-induced breakdown, we do not discuss it in further detail here.

For the Type-III and Type-IV regimes, the adiabatic levels are connected at the knot point. A natural question arises: when a quantum state passes adiabatically through this knot, does it continue to follow one of the adiabatic branches, or does it split into a superposition of the two? Our calculations show that neither of these conventional scenarios occurs.

As illustrated in Fig. \ref{fig:fig4}(c), which corresponds to the Type-III regime (positive slope at the knot), the dynamical level closely tracks the lower adiabatic branch before reaching the knot. However, after traversing the knot, the state does not remain on a single adiabatic branch, nor does it split into a coherent superposition. Instead, the dynamical level departs from both adiabatic curves, exhibiting a trajectory that is not captured by any instantaneous eigenstate. A similar departure is observed in the Type-IV regime [Fig. \ref{fig:fig4}(d)].

To further elucidate this behavior, we examine in Fig. \ref{fig:fig4}(d) not only the evolution starting from the lower branch (circles) but also that initiated from the upper adiabatic branch (triangles). Surprisingly, after passing through the four-real-root region, the two dynamical trajectories become indistinguishable---regardless of their initial condition, they converge to the same dynamical path. \emph{This indistinguishability of initially distinct adiabatic trajectories} is a hallmark of the nonlinear-coupling-induced restructuring of phase space and stands in stark contrast to the behavior of systems with only on-site nonlinearity.

In the following text, we provide a detailed analysis of the underlying physics. Using a fixed-point description of the equivalent classical Hamiltonian, we show that the convergence of the two dynamical trajectories arises from \emph{the unique black-hole-like fixed point of the system, which effectively erases the memory of the initial adiabatic branch}. This mechanism, which is unique to nonlinear coupling, explains both the finite adiabatic tunneling probability and the unconventional dynamical merging observed in Fig. \ref{fig:fig4}(c) and \ref{fig:fig4}(d).

\section{Nonlinear quantum adiabatic theorem}\label{AdiabaticTheorem}
\subsection{The black-hole-like fixed point}
Based on the given matrix equation \ref{eq:ham}, we introduce the population difference \(s = |b|^2 - |a|^2\) and the phase difference \(\theta = \theta_b - \theta_a\). Employing the probability conservation condition \(|a|^2 + |b|^2 = 1\), we derive the following canonical equations:
\begin{subequations} \label{eq:CanonicalEquations}
\begin{align}
&\dot{s} = -\left(2\alpha + \beta(1-s)\right) \sqrt{1-s^2} \sin \theta, \label{eq:CanonicalEquationsA}\\
&\dot{\theta} = 2\gamma + \left(2\alpha + \beta(1-s)\right) \dfrac{s}{\sqrt{1-s^2}} \cos \theta. \label{eq:CanonicalEquationsB}
\end{align}
\end{subequations}
These equations \eqref{eq:CanonicalEquationsA} and \eqref{eq:CanonicalEquationsB} constitute a classical Hamiltonian system with respect to the non-canonical Poisson bracket \(\{ \theta, s \} = 2\alpha + \beta(1-s)\), satisfying
\begin{equation}\label{eq:CanonicalEquation}
\dot{s} = \{ s, H_c \} = -\{ \theta, s \} \frac{\partial H_c}{\partial \theta}, \quad \dot{\theta} = \{ \theta, H_c \} = \{ \theta, s \} \frac{\partial H_c}{\partial s}.
\end{equation}
The corresponding classical Hamiltonian is obtained as:
\begin{equation}\label{eq:ClassicalHamiltonian}
H_c(s, \theta) = -\sqrt{1-s^2} \cos \theta - \frac{2\gamma}{\beta} \ln\left|2\alpha + \beta(1-s)\right|.
\end{equation}

If one instead adopts the standard Poisson bracket \(\{ \theta, s \} = 1\), the system does not assume a standard Hamiltonian form for \(\beta \neq 0\). Only in the linear limit \(\beta = 0\) does the Hamiltonian reduce to \cite{Anderson1963,PhysRevB.40.2158,PhysRevLett.79.4950} \(H_c = 2\gamma s - 2\alpha \sqrt{1-s^2} \cos \theta\), which then satisfies the standard canonical relations \(\dot{s} = -\partial H_c/\partial \theta\) and \(\dot{\theta} = \partial H_c/\partial s\).

For a system described by the canonical equations \eqref{eq:CanonicalEquationsA} and \eqref{eq:CanonicalEquationsB}, the fixed points are determined by the conditions \(\dot{s}=0\) and \(\dot{\theta}=0\). (\textbf{i}) For \(\gamma = 0\), there are always two isolated fixed points: \((s=0, \theta=0)\) and \((s=0, \theta=\pi)\), and one continuous curve of fixed points for $\beta<-1$: \(s_{\rm hole} = 1 + 2\alpha/\beta\) with \(\theta\) arbitrary. It is a unique fixed point induced by nonlinear coupling, which has never been discovered in on-site nonlinear systems. (it is named as the black-hole-like fixed point in the following text). (\textbf{ii}) For \(\gamma \neq 0\), the fixed point satisfies the following equation
\begin{equation}\label{eq:FixedPointEquation}
\bigl(2\alpha + \beta(1-s)\bigr) \frac{s}{\sqrt{1-s^2}} = -\frac{2\gamma}{\sigma},
\end{equation}
where \(\sigma = \cos \theta = \pm 1\) (\(\sigma = 1\) for \(\theta=0\); \(\sigma = -1\) for \(\theta=\pi\)). Defining \(h(s) = \bigl(2\alpha + \beta(1-s)\bigr) s / \sqrt{1-s^2}\), this becomes \(h(s) = -2\gamma/\sigma\). For each \(\sigma\), the equation may have 0, 1, or 2 real solutions for \(s \in (-1,1)\), each corresponding to a fixed point \((\theta, s)\). The number and nature of fixed points depend sensitively on \(\alpha, \beta, \gamma\), necessitating a case-by-case analysis.
\begin{figure}[tb]
\centering
\includegraphics[width=\columnwidth]{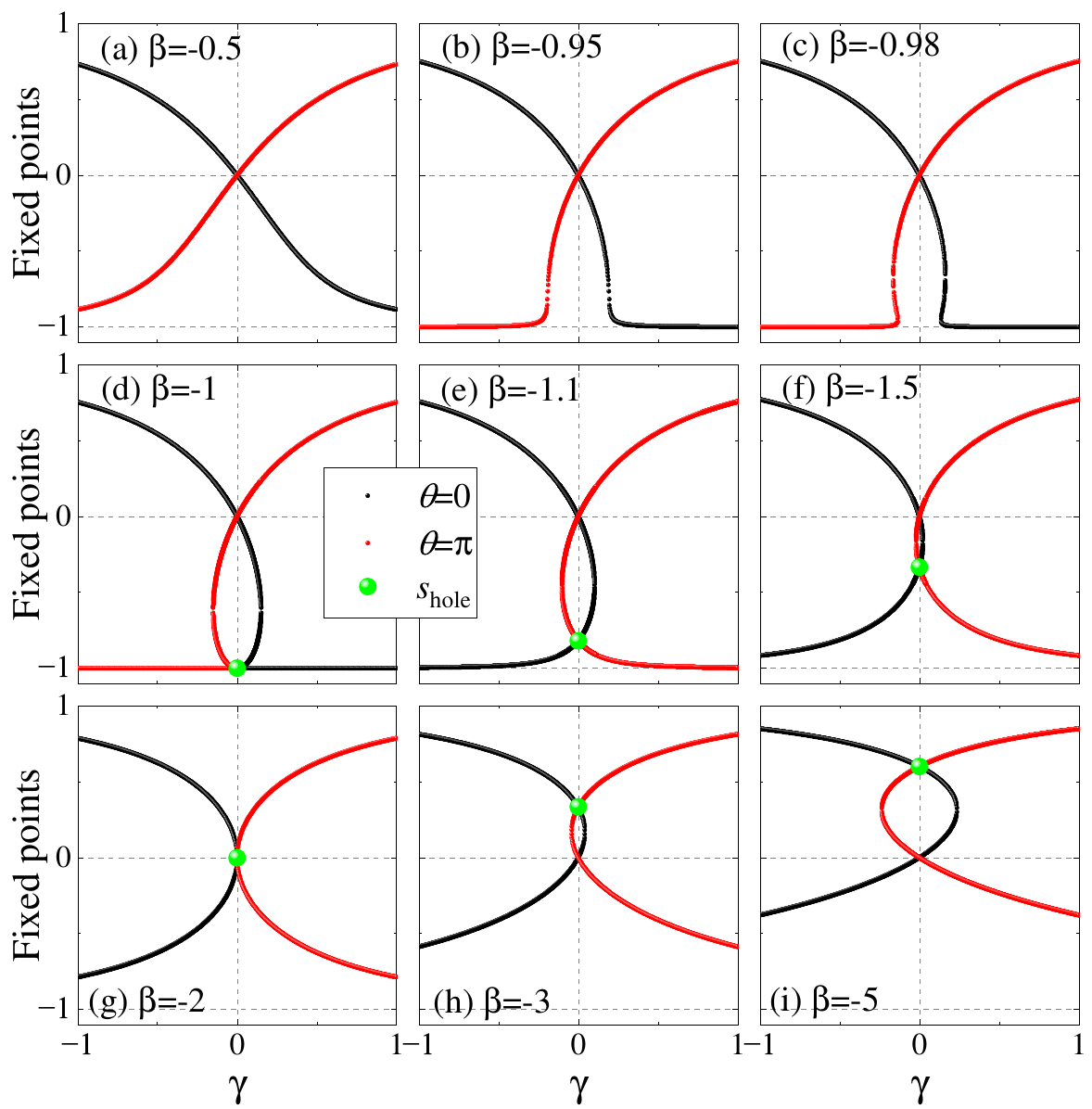}
\caption{Fixed points \(s_{\rm f}\) as a function of bias \(\gamma\) for different values of the nonlinear coupling strength \(\beta\). The nonlinear coupling qualitatively modifies the fixed-point topology and count. Green symbols mark the black-hole-like fixed points. }
\label{fig:fig5}
\end{figure}

Figure \ref{fig:fig5} details how the structure and number of fixed points evolve with the bias \(\gamma\) for different values of the nonlinear coupling strength \(\beta\). In the regime \(\beta > -0.9575\) [Figs. \ref{fig:fig5}(a) and \ref{fig:fig5}(b)], only two fixed points exist throughout the entire range of \(\gamma\). When \(-1 < \beta < -0.9575\), a window in \(\gamma\) opens in which the number of fixed points increases to four. A similar four-fixed-point window appears again for \(\beta < -1\). The figure clearly illustrates that the number of fixed points located on the \(\theta = 0\) and \(\theta = \pi\) branches can be effectively controlled by the nonlinear coupling strength.

In fact, the variation in the number of fixed points follows exactly the same pattern as the change in the number of real roots of the adiabatic energy equation shown in Fig. \ref{fig:energylevels}. This correspondence reflects the deep connection between the fixed-point topology and the real-solution structure of the adiabatic spectrum. In the following, we carry out a detailed analysis of the fixed-point equation \eqref{eq:FixedPointEquation} to derive the exact conditions for these transitions as shown in Fig. \ref{fig:energyphase}.

The fixed-point equation \eqref{eq:FixedPointEquation} yields four distinct real roots if and only if the following condition is satisfied for \(\beta < -1\) (see Appendix \ref{SecAppendix1} for a detailed derivation):
\begin{equation}\label{eq:boundary1}
|\gamma| < \gamma_c(\beta)=f(\beta) = -\frac{(\beta+2)^2}{8\beta \sqrt{1 - s_v^2}},
\end{equation}
where \(s_v=1/2+1/\beta\) .

For the regime \(-1 < \beta < -0.9575\), the region supporting four distinct real roots is bounded by two separate curves, whose asymptotic expansions are (derived in Appendix \ref{SecAppendix2})
\begin{align}\label{eq:boundary2}
\gamma_{c1}(\beta) &=g_1(\beta) = \sqrt{\eta}\Bigl(1 - 2\eta - 6\eta^2 - 50\eta^3 + O(\eta^4)\Bigr), \\[4pt]
\gamma_{c2}(\beta) &=g_2(\beta)= \kappa_0 + \kappa_1 \eta + \kappa_2 \eta^2 + \kappa_3 \eta^3 + O(\eta^4),
\end{align}
with \(\eta = \beta + 1\) and the coefficients \(\kappa_0\approx 0.150,\quad \kappa_1 \approx 0.34, \quad \kappa_2 \approx 1.5\), and \(\kappa_3 \approx 5.0\) .
These analytical boundaries are plotted in Fig. \ref{fig:energyphase} together with the numerically obtained phase regions; the two show excellent agreement.

\subsection{Nonlinear quantum adiabatic theorem}
Our calculations reveal that, in the nonlinear-coupling-induced LZ tunneling process, the adiabatic evolution is governed by a black-hole-like fixed point. This object plays a pivotal role in shaping the system¡¯s dynamical pathway and is responsible for the singular behavior observed near the critical point. In the following, we elucidate its decisive influence on adiabatic dynamics and, building on this insight, derive an analytical expression for the adiabatic tunneling probability.
\begin{figure}[tb]
\centering
\includegraphics[width=\columnwidth]{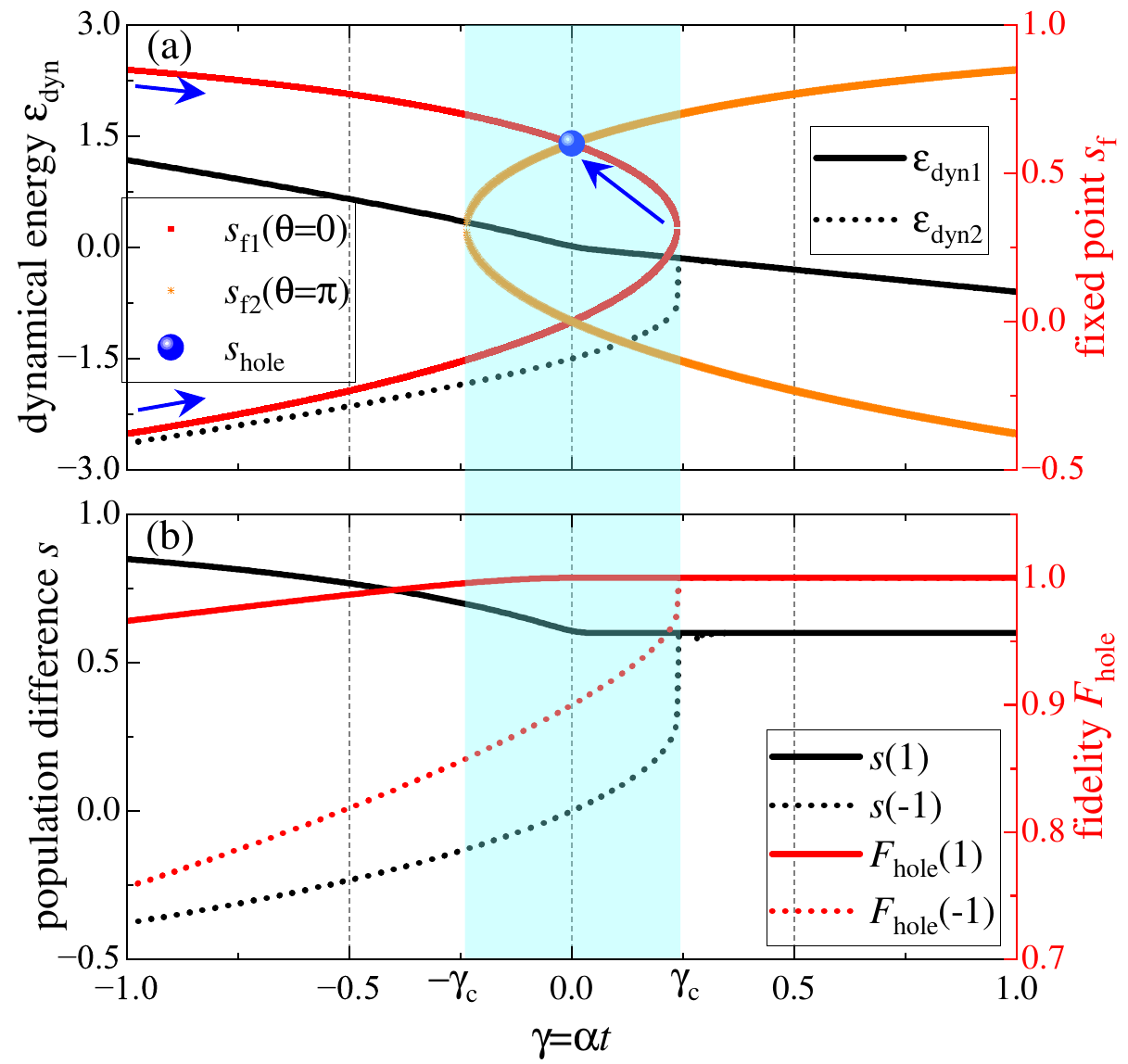}
\caption{The critical role of the black-hole-like fixed point in adiabatic quantum dynamics.
This figure employs dual ordinate axes.
(a) Dynamical energy \(\varepsilon_{\mathrm{dyn}}(\gamma)\) (left ordinate) and the fixed points \(s_{\mathrm{f}}\) (right ordinate) as functions of the level bias \(\gamma = \alpha t\) in the adiabatic limit (i.e., \(v = 0.0001\)).
Blue dots mark the black-hole-like fixed point $s_{\rm hole}$.
The two dynamical energy curves correspond to two distinct initial states:
\((a(\gamma\to -\infty), b(\gamma\to -\infty))^{\mathrm{T}} = (1,0)^{\mathrm{T}}\) (black dashed line) and
\((a(\gamma\to -\infty), b(\gamma\to -\infty))^{\mathrm{T}} = (0,1)^{\mathrm{T}}\) (black solid line).
Blue arrows indicate the direction of quantum state evolution.
(b) Population difference \(s\) (left ordinate) and the adiabatic fidelity \(F_{\mathrm{hole}}\) (right ordinate) versus the level bias \(\gamma = \alpha t\) in the adiabatic limit (i.e., \(v = 0.0001\)).
The two population difference curves correspond to initial population imbalances:
\(s(\gamma\to -\infty) = -1\) (black dashed line) and
\(s(\gamma\to -\infty) = 1\) (black solid line).
The two adiabatic fidelity curves correspond to the same initial states as in (a):
\((a,b)^{\mathrm{T}} = (1,0)^{\mathrm{T}}\) (red dashed line) and
\((a,b)^{\mathrm{T}} = (0,1)^{\mathrm{T}}\) (red solid line).
}
\label{fig:fig6}
\end{figure}

Figure \ref{fig:fig6}(a) displays the dynamical energy \(\varepsilon_{\mathrm{dyn}}(\gamma)\) (left ordinate) together with the fixed points \(s_{\mathrm{f}}\) (right ordinate) as functions of the bias \(\gamma = \alpha t\) in the adiabatic limit (exemplified by \(v = 0.0001\)). The fixed-point branches clearly delineate the window within which four real energy levels coexist---shaded region \((-\gamma_c, \gamma_c)\). Two dynamical energy trajectories, corresponding to two distinct initial conditions \( (a,b)^{\mathrm{T}} = (1,0)^{\mathrm{T}} \) (black dashed) and \( (0,1)^{\mathrm{T}} \) (black solid), are shown. Remarkably, after traversing the shaded region the two trajectories collapse onto each other, indicating that all memory of the initial state is completely erased by passage through this region.

Figure \ref{fig:fig6}(b) shows the corresponding population difference \(s\) (left ordinate) versus \(\gamma\) under the same adiabatic conditions. The two curves correspond to initial imbalances \(s(\gamma\to -\infty) = -1\) (black dashed) and \(s(\gamma\to -\infty) = 1\) (black solid). Upon adiabatic crossing of the shaded region, they too become indistinguishable. A striking observation is that the curve starting from \(s = 1\) encounters the black-hole-like fixed point \(s_{\mathrm{hole}}\) and thereafter locks to a constant value---precisely the value of \(s_{\mathrm{hole}}\) itself. The curve starting from \(s = -1\) reaches the boundary \(\gamma_c\); instead of continuing adiabatically along the \(\theta = \pi\) branch, it is captured by the basin of attraction of \(s_{\mathrm{hole}}\) and subsequently remains at the same constant. This behavior suggests a bold conjecture: whenever a black-hole-like fixed point lies on the adiabatic path, the asymptotic state after the passage is determined solely by this fixed point, completely independent of the initial condition.

To substantiate this conjecture, we compute the adiabatic fidelity \(F_{\mathrm{hole}}(\gamma)\), defined as the projection of the evolving quantum state onto the black-hole-like fixed point in the adiabatic limit: $F_{\mathrm{hole}}(\gamma) = \lim_{v\to 0} \bigl| \langle \psi(\gamma) | \psi_{\mathrm{hole}}(\gamma) \rangle \bigr|^2.$
The results are overlaid in Fig. \ref{fig:fig6}(b) (red curves). For the initial state \((a,b)^{\mathrm{T}} = (0,1)^{\mathrm{T}}\) (red solid line), \(F_{\mathrm{hole}}\) rises to unity immediately after the system passes \(s_{\mathrm{hole}}\). For the initial state \((1,0)^{\mathrm{T}}\) (red dashed line), \(F_{\mathrm{hole}}\) jumps to unity upon crossing \(\gamma_c\). Hence, in all cases the system becomes perfectly projected onto \(s_{\mathrm{hole}}\) once the shaded region is exited---a direct verification of our conjecture.

The adiabatic-following analysis of Fig. \ref{fig:fig4}, combined with the fixed-point perspective of Fig. \ref{fig:fig6}(a), unveils a previously unrecognized adiabatic transport scenario. As schematized in Fig. \ref{fig:fig6}(a), the upper fixed-point branch \(s_{\mathrm{f1}}\) evolves adiabatically with \(\gamma\) until it collides with the black-hole-like fixed point \(s_{\mathrm{hole}}\). Upon collision, it is trapped at \(s_{\mathrm{hole}}\) and ceases to evolve further; all information about its prior evolution---initial condition and path---is completely erased. Meanwhile, the lower fixed-point branch \(s_{\mathrm{f1}}\) continues to follow the adiabatic trajectory until it reaches the boundary \(\gamma_c\). At this point no continuous adiabatic path exists along the \(\theta =0\) manifold. Instead of jumping to the other branch (\(\theta =\pi\)), the system is drawn into the basin of attraction of \(s_{\mathrm{hole}}\) and asymptotically settles there. Again, any trace of its origin is wiped out. Owing to this all-absorbing nature, we term this special fixed point a black-hole-like fixed point.

The presence of the black-hole-like fixed point provides the physical mechanism underlying the residual adiabatic tunneling probability. By projecting the asymptotic state onto the instantaneous eigenbasis, and exploiting the fact that after passing the critical region the system becomes pinned to \(s_{\mathrm{hole}}\), one can directly read off the final population distribution. Using the relation \(s = |b|^2 - |a|^2\) together with probability conservation \(|a|^2 + |b|^2 = 1\), the adiabatic tunneling probability---defined as the population transferred to the upper adiabatic branch---follows simply as
\begin{equation} \label{eq:pad}
p_{\mathrm{ad}} = -\frac{\alpha}{\beta} \qquad (\beta < -1).
\end{equation}
Remarkably, this compact expression depends only on the ratio of the coupling parameters and holds across the entire supercritical regime. The derivation, which is based solely on the fixed-point coordinates and requires no detailed knowledge of the full time evolution, underscores the predictive power of the black-hole-like fixed point picture. As shown in Fig. \ref{fig:fig3}(d), the analytical result of Eq. (\ref{eq:pad}) agrees perfectly with the numerical simulations for all \(\beta < -1\), confirming both the validity of the underlying assumptions and the essential role played by the black-hole-like fixed point in governing adiabatic dynamics.

\section{Conclusion}\label{Conclusion}
We have developed a comprehensive theory of LZ tunneling in a two-level system with amplitude-dependent nonlinear coupling---a distinct mechanism from conventional on-site nonlinearity. Through a detailed fixed-point analysis, we have fully characterized the adiabatic energy landscape and established the exact conditions for the emergence of swallowtail and twisted-knotted structures. The phase diagram in the \((\beta,~\gamma)\) plane is analytically mapped out, and the critical boundaries separating two-level from four-level regimes are derived in closed form, showing perfect agreement with numerical simulations.

A central discovery of our work is the identification of a black-hole-like fixed point that governs adiabatic evolution in the supercritical regime. This fixed point acts as a universal attractor: regardless of the initial state, the dynamical trajectory converges to it upon traversing the critical region. All information about the initial condition and the detailed path is completely erased---an effect without analogue in systems with only on-site nonlinearity. The asymptotic state is uniquely determined by the coordinates of this fixed point, yielding a remarkably simple expression for the adiabatic tunneling probability, \(p_{\mathrm{ad}} = \alpha / \beta\) (\(\beta < -1\)), which we verified numerically.

Our results reveal that nonlinear coupling induces a topological reorganization of phase space that fundamentally alters the adiabatic theorem. Even in the limit of an infinitely slow sweep, a finite tunneling probability persists, and dynamical trajectories emanating from distinct adiabatic branches become indistinguishable after passing through the four-level window. This 'indistinguishability' reflects the breakdown of adiabatic following and provides a striking illustration of how nonlinear interactions can override the constraints of linear adiabatic dynamics.

The framework established here is generic. It applies to any driven two-mode system with state-dependent coupling---ranging from acoustic resonators and photonic waveguides to ultracold atoms in optical lattices---and can be extended to time-modulated, non-Hermitian, or multi-site configurations. Our findings open avenues for the deliberate engineering of adiabatic passage, the design of information-erasing protocols, and the control of multistable dynamics in nonlinear wave systems. More broadly, they establish nonlinear coupling as a powerful and versatile tool for shaping quantum and classical adiabaticity, with potential applications in topological state preparation, atomtronics, and programmable analog quantum simulation.

\section*{Acknowledgments}
This work was supported by the National Natural Science Foundation of China (Contracts No. 12005173 and No. 12365004), the Natural Science Foundation of Gansu Province (Contract No. 20JR10RA082).

\section*{Data availability}
The data that support the findings of this article are not publicly available. The data are available from the authors upon reasonable request.

\appendix
\renewcommand{\thefigure}{A\arabic{figure}}
\renewcommand{\theequation}{A\arabic{equation}}
\setcounter{figure}{0}
\setcounter{equation}{0}

\section{Boundary conditions for the number of real fixed points}\label{SecAppendix}

\subsection{Boundary for the regime \((\beta < -1)\)}\label{SecAppendix1}

After setting \(\alpha = 1\), the fixed-point equation \eqref{eq:FixedPointEquation} reduces to
\[
2|\gamma| \sqrt{1 - s^{2}} = \bigl| s\bigl[\beta s - (\beta+2)\bigr] \bigr|,
\qquad s\in[-1,1].
\]
Define \(g(s)=\beta s^{2}-(\beta+2)s\). The equation then describes the intersection of the curve \(L(s)=2|\gamma|\sqrt{1-s^{2}}\) with \(|g(s)|\). By analysing the shapes of these two functions on the interval \([-1,1]\) we obtain the condition for the number of real roots. Let
\[
M_{1}=|g(s_{v})| = g(s_{v}) = -\frac{(\beta+2)^{2}}{4\beta}
\qquad (\text{positive because }\beta<0),
\]
and \(L(s_{v})=2|\gamma|\sqrt{1-s_{v}^{2}}\) with \(s_{v}=1/2+1/\beta\). Four distinct real roots appear if and only if
\[
|\gamma| \;<\; \gamma_{c}(\beta)\equiv
-\frac{(\beta+2)^{2}}{8\beta\sqrt{1-s_{v}^{2}}},
\]
where the square root in the denominator is always positive.

When \(|\gamma|=\gamma_{c}(\beta)\) the equation has three distinct real roots (one simple and one double root); for \(|\gamma|>\gamma_{c}(\beta)\) only two distinct real roots survive.

\subsection{Boundary for the regime \((-1<\beta<-0.9575)\)}\label{SecAppendix2}

For \(\beta\in(-1,-0.9575)\) the number of real roots of (7) is governed by two critical values \(\gamma_{c1}(\beta)\) and \(\gamma_{c2}(\beta)\) with \(0<\gamma_{c1}<\gamma_{c2}\). These thresholds correspond to the appearance of a double root (tangency) in the interval \(s\in(-1,0)\), given by the system
\[
\begin{cases}
\beta s^{2}-(\beta+2)s = 2|\gamma|\sqrt{1-s^{2}},\\[4pt]
2\beta s-(\beta+2) = -\dfrac{2|\gamma|\,s}{\sqrt{1-s^{2}}}.
\end{cases}
\]

Eliminating \(|\gamma|\) yields a cubic equation for \(s\),
\[
\beta s^{3}-2\beta s+(\beta+2)=0,
\]
which possesses two real roots \(s_{1},s_{2}\in(-1,0)\). Substituting them back gives \(\gamma_{c1}=|\gamma(s_{1})|\) and \(\gamma_{c2}=|\gamma(s_{2})|\).

A perturbative expansion near \(\beta=-1\) (writing \(\beta=-1+\eta\) with small \(\eta>0\)) provides explicit approximations.

For the lower boundary \(\gamma_{c1}\) (associated with a tangency near \(s\approx-1\)):
\[
\gamma_{c1}= \sqrt{\eta}\,
\Bigl(1-2\eta-6\eta^{2}-50\eta^{3}
      -\tfrac{6927}{4}\eta^{4}+O(\eta^{5})\Bigr),\qquad
\eta=\beta+1.
\]

For the upper boundary \(\gamma_{c2}\) (associated with \(s_{2}\approx\varphi\equiv\frac{1-\sqrt5}{2}\approx-0.618\)):
\[
\gamma_{c2}= \kappa_{0}+\kappa_{1}\eta+\kappa_{2}\eta^{2}
            +\kappa_{3}\eta^{3}+O(\eta^{4}),
\]
with
\[
\kappa_{0}= \frac{(2\varphi+1)\sqrt{-\varphi}}{2\varphi}\approx0.150,\qquad
\kappa_{1}\approx0.34,\;\; \kappa_{2}\approx1.5,\;\; \kappa_{3}\approx5.0 .
\]
Higher-order coefficients can be obtained by continuing the expansion; the above suffices for \(\eta\ll1\).

Summary of the root count in the regime \((-1<\beta<-0.9575)\):
\begin{itemize}
    \item Two real roots when \(|\gamma|<\gamma_{c1}\) or \(|\gamma|>\gamma_{c2}\).
    \item Four real roots when \(\gamma_{c1}<|\gamma|<\gamma_{c2}\).
    \item Three real roots (one double root) when \(|\gamma|=\gamma_{c1}\) or \(|\gamma|=\gamma_{c2}\).
\end{itemize}

As \(\eta\) increases (i.e. \(\beta\) grows from \(-1\)), \(\gamma_{c1}\) and \(\gamma_{c2}\) approach each other; they merge at \(\beta\approx-0.9575\), beyond which the four-root region disappears.

\end{document}